\input{aipcheck}

\documentclass[
    ,final            
  ]
  {aipproc}

\layoutstyle{6x9}

\def\GS{GS~1826$-$238~}
\def\SAX{{\em BeppoSAX}~}
\def\XTE{{\em RXTE}~}
\def\XSPEC{{\sc Xspec}~}
\def\ergcms{{\rm erg~cm$^{-2}$s$^{-1}$}~}
\def\CompTB{{\sc comptb}~}
\def\CompTT{{\sc comptt}~}

\def\aap{\emph{A\&A}~}
\def\apj{\emph{ApJ}~}

\begin{document}

\title{
Two separate spectral substates within Low-Hard States of NS-LMXBs}



\author{M. Cocchi}{
address={INAF, IASF sez. di Roma, via Fosso del Cavaliere, 100, 00133 Roma (Italy)}
}

\begin{abstract}
The whole \SAX archive was screened for observations of Neutron Star (NS) Low-Mass X-ray Binaries,
in order to characterise the X-ray properties of
this class of sources when observed in a Low Hard State (LHS). 
A total of 12 sources in the sample exhibited LHS spectra; half of the
objects showed a \textit{canonical} LHS spectrum, in which two separate populations
of soft blackbody-distributed photons emerge. One, or possibly both
the populations, is Comptonized by a hot ($kT_{e} > 20$ keV) electron plasma cloud.
For the remaining 6 sources just a single population is sufficient to account for both the
directly observed blackbody and the Comptonized radiation.

The luminosities of the NS with canonical two-population spectra are
found to be systematically higher than the ones of the sources
showing the atypical, single photon population spectra.
This suggests the possibility of two separate subclasses of LHS
spectra. The transition from a subclass to the other is likely related to the
accretion rate, which drives the presence of a hotter X-ray photon
population in the close vicinity of the NS, or possibly a stronger
thermal gradient in the inner, X-ray active, accretion disk.
\end{abstract}

\maketitle


\section{The canonical Low-Hard State spectra}
Neutron Stars (NS) in Low Mass X-ray Binaries (LMXBs) are typically observed in two main spectral states: High/Soft State (HSS) and Low/Hard State (LHS).
Both HSS and LHS spectra are generally described by two (main) components: a Comptonized radiation of soft seed photons, dominating the high energy output, and a soft, blackbody-like, emission, detected at lower energies. 
The temperature of the Comptonizing electron plasma is very different in the two states ($\sim 3$ keV in HSS and $\sim 15-25$ keV in LHS) and characterizes the wide band spectral shape: a big soft X-ray bump with no or poor high energy emission in HSS, a small soft bump and evident hard-X extension (up to $\sim 200$ keV) in LHS.

As the temperature of the blackbody component is generally different from the one of the Compton-modified seed photons, both the states are characterised by two well separated, blackbody-like, soft X-ray photon populations.  Depending on the temperatures of these populations, two classical models properly fit the data: the \textit{Eastern} model \cite{Mitsuda}, with a cold ($kT \sim 0.5$ keV), disk-like, directly observed photon population and hotter ($kT=1-2$ keV) Comptonized population; and the \textit{Western} model \cite{White}, with hot, NS originating, blackbody and colder Comptonized seed photons.
At least for LHS spectra, Eastern-like models look pereferable (e.g. \cite{Guainazzi} for a \SAX application) also because this kind of models deal with very compact X-ray emitting regions, which is required by the quick variability of NS-LMXBs.  Up to now, the multicolour nature of the directly observed blackbody is not confirmed by the wide band data, as pure blackbody is also sufficient to model the disk photon population.

\section{Analysis of the \SAX NS-LMXB sample}
The \SAX on-line archive of ASI Science Data Centre (ASDC) provides standard analysis products (spectra, lightcurves) for each performed observation\footnote{\url{http://www.asdc.asi.it/bepposax}}.
Joint MECS, LECS and PDS spectra, deconvolved by the officially released \SAX response matrices and ancillary files, can be thus analysed; for this work, the \XSPEC fitting tool (version 12) was used. HPGSPC spectra are generally not available as ASDC products, so they are not considered for this preliminary \SAX overview of NS-LMXBs.

The sample includes as many as 22 sources, 10 of which were observed in HSS only. Two of the 12 sources with LHS spectra (namely SLX 1735-269 and 4U 1745$-$203 in NGC 6440) were also detected in HSS.

For the analysis of the Compton-dominated LHS spectra the recently developed model \CompTB\footnote{\url{http://heasarc.nasa.gov/docs/xanadu/xspec/newmodels.html}} \cite{Farinelli} was adopted.
\CompTB is an up-to-date Comptonization model in a diffusion regime including in a self-consistent way the seed photons emission together with its Comptonization, be the latter either purely thermal or mixed thermal plus dynamical (bulk Comptonization).
As no detectable bulk effects are expected in LHS \cite{Paizis}, here \CompTB is always used in a purely thermal fashion, similarly to the well-known model \CompTT \cite{Titarchuk}.
In \eqref{CompTB} the model is outlined: the Comptonized emission of an empirically modified blackbody (BB) 
is given by the convolution term $BB*G$, where G is a general Green function.
\begin{equation}
f(E)=\frac{N}{1+A}\left[BB+A(BB*G)\right] ~~~~~~~  {\rm where} ~~~~~  BB(x)\propto\frac{x^{\gamma}}{e^{x}-1}  \label{CompTB}
\end{equation}
In thermal mode the free parameters of $BB*G$ are the energy index of the G function, 
$\alpha$ ($\Gamma = \alpha +1$) and the electron plasma temperature, $kT_{e}$.
The modified blackbody $BB = BB(E/k{\rm T}_{s})$ has $k{\rm T}_{s}$, the index $\gamma$ and normalization as main parameters;
in this paper always $\gamma=3$, i.e. pure blackbody (the model's default). 
The illumination factor A, parametrized as log(A), indicates the output contribution of the directly observed seed blackbody radiation with respect to its Comptonization.

\begin{figure}
  \includegraphics[height=0.44\textheight]{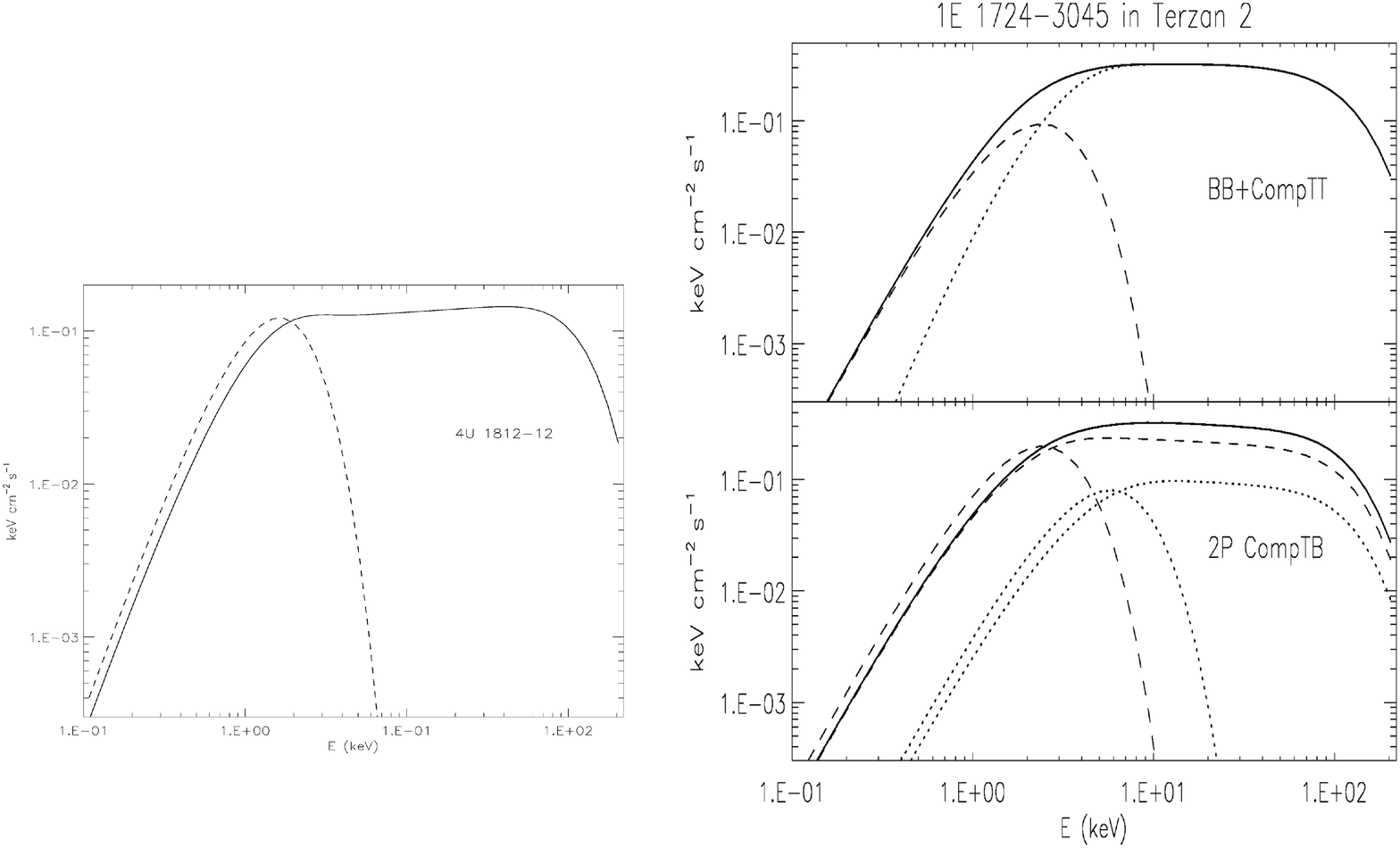}
  \caption{
           \textbf{Left:} The best \CompTB fit model for 4U 1812$-$12: the blackbody seed photon population is also shown (dashed line).
           \textbf{Right:} \textit{lower panel}, the two-seeded Comptonized best fit model for Terzan 2: the two \CompTB components (disk, dashed line, and NS-TL, dotted) are also shown, along with their original blackbody seed photon populations.~~\textit{Upper panel}, the canonical blackbody + \CompTT best fit model of the same data.
}
\end{figure}

\section{Data analysis results: LHS spectra substates}
A canonical, two-population model was applied to the sample of 12 LHS sources.
As e.g. in \cite{Guainazzi} the spectral components were: 1) Comptonization of NS or transition layer (TL) soft photons, modeled by \CompTT (or, similarly, \CompTB); 
2) direct disk photon emission, approximated by a pure blackbody.
Interstellar H absorption, modeled by {\sc wabs} in \XSPEC, was also included.
Even though the fits were acceptable, for 6 sources the temperatures of the two populations were found to be very similar, sometimes overlapping within the statistical errors. This indicates that, in such cases, the data do not require the hypothesis of two distinct seed photon populations. Therefore in this \SAX sample two kind of LHS spectra can be distinguished:
\begin{itemize}
\item
 \textbf{Single-population spectra (1P).}
      The best fit is provided by a {\sc wabs}*\CompTB model, with a single blackbody photon population by definition.  Part of the blackbody is directly observed, the rest is Comptonized by a hot ($kT_{e} > 20$ keV) electron plasma. 
      A good example of a 1P LHS spectrum, the source 4U 1812$-$12, is shown in Fig.1 (\textit{left}).
      The main best fit parameters are: $n_{\rm H}= 1.5\pm 0.1 \times 10^{22} {\rm cm}^{-2}$; $kT_{s}=0.41\pm 0.04$ keV; $\alpha=0.94\pm0.03$;  $kT_{e}=25\pm 3$ keV; $logA=0.7\pm 0.1$; $\chi^{2}_{r}=0.84$ (197 d.o.f.).
\item
 \textbf{Two-population spectra (2P)}.
      In this case two blackbody populations are actually needed, and a traditional {\sc wabs}*({\sc bb}+\CompTT) model applies.
      As in \cite{Cocchi2}, also a model assuming both the soft populations to be Comptonized by the same electron cloud well fits to the \SAX data, and in some cases seems to be slightly preferable. This model, implemented by {\sc wabs}*(\CompTB+\CompTB) in \XSPEC with frozen logA=8 (no direct BB), yelds plasma temperatures in the 18$-$25 keV range and seed photon colour temperatures in the range 0.4--0.6 keV and 1.4--2.2 keV for the disk and the NS/TL populations, respectively.

      An example of a 2P spectrum, Terzan 2, is also shown in Fig.1 (\textit{right)}.
The main best fit parameters for Terzan 2 are: $n_{\rm H}= 0.84\pm 0.05 \times 10^{22} {\rm cm}^{-2}$; $kT_{s}(disk)=0.62\pm 0.04$ keV; $kT_{s}(NS/TL)=1.5\pm 0.3$ keV; $\alpha = 1.07\pm 0.03$; $kT_{e}= 25\pm 1$ keV; $\chi^{2}_{r}= 1.17$ (200 d.o.f.).
\end{itemize}


\begin{table}
\begin{tabular}{lccccl}
\hline
  \tablehead{1}{r}{b}{Source name}
  & \tablehead{1}{r}{b}{\SAX\\pointings\tablenote{only the analysed ones}}
  & \tablehead{1}{r}{b}{2P type\\bol. flux\tablenote{unabsorbed 0.1$-$200 keV flux in $10^{-10}$ \ergcms}}
  & \tablehead{1}{r}{b}{1P type\\bol. flux}
  & \tablehead{1}{r}{b}{UCX\\candidate \cite{Zand}}
  & \tablehead{1}{r}{b}{Globular\\Cluster}   \\
\hline
SAX 1712$-$373 & 1 & - &  3 & yes & \\
SLX 1735$-$269 & 1 & - &  5 & yes & \\
SLX 1737$-$282 & 1 & - &  2 & yes & \\
SAX 1810$-$260 & 1 & - &  4 &     & \\
4U 1812$-$12   & 1 & - & 12 & yes & \\
4U 1850$-$087  & 1 & - &  3 & yes & NGC 6712 \\
\hline
4U 0614$+$91   & 2 & 24, 19 & - & yes  & \\
4U 1702$-$429  & 1 & 18 & - &    & \\
1E 1724$-$307  & 1 & 20 & - &    & Terzan 2 \\
4U 1745$-$203  & 1 & 14 & - &    & NGC 6440 \\
SAX 1750$-$290  & 1 & 22 & - &   &  \\
GS 1826$-$238  & 6 & 19$-$23 & - &   & \\
\hline
\end{tabular}
\caption{The analysed \SAX NS-LMXB sample (only the sources detected in LHS are mentioned).}
\label{tab1}
\end{table}

\section{Discussion}
Assuming there are two subclasses of LHS, it is natural to look for unifying parameters able to physically explain their spectral differences.
Table 1 shows how 1P LHS are systematically fainter than the 2P ones: the measured unabsorbed bolometric fluxes are $2-12\times 10^{-10}$ \ergcms for 1P spectra and $14-24\times 10^{-10}$ \ergcms for the 2Ps. Even neglecting the obvious distance bias (e.g. 4U 1812$-$12 is only $\sim 4$ kpc away \cite{Cocchi1}) and assuming a standard distance of 10 kpc, 1P luminosities are all below the threshold of $\sim 10^{37}$ erg/s, while 2P are above that critical value, corresponding to a $\sim 3\%$ Eddington luminosity.
So the accretion rate ($\dot m$) can drive the differences among the two subclasses: 
at low $\dot m$, relatively cold disk photons account for most of the soft population. Though they are likely multicolour-distributed, pure blackbody suffices to fit the wide band data. Part of this disk photon population is Comptonized to high energies by the hot electron cloud.
As $\dot m$ increases, the contributions of NS and TL hotter populations become important, i.e. a second photon population sets in. Or, alternatevely, a stronger temperature gradient is formed between the outer and the inner (close to NS) disk: in the fits such gradient is approximated by two photon populations.

1P spectra look \textit{intrinsically} one-population, i.e. it's not a counting statistics bias: assuming two populations yields very close BB temperatures and/or fit insensitivity to $kT_{s}$ (e.g. this is true for the relatively bright 4U 1812$-$12).
Most of the sources with 1P spectra are Ultra-Compact (UCXB) candidates \cite{Zand}; this is not surprising, as UCX candidates are selected mainly because of their low persistent luminosity. Though the nature of UCXBs is probably different from that of the brighter NS-LMXBs, their spectral characteristics look nevertheless related to the accretion rate: UCXB candidates are also detected in brighter, softer states (e.g. SLX 1735$-$269 and 4U 1820$-$30 in this \SAX sample) or in a 2P substate (4U 0614$+$91).

As already shown, 2P spectra are best-fit by both a traditional uncovered-disk + covered-NS/TL Eastern-like model and a completely Comptonized two seeded model: there is not clear evidence for one soft photon population to be uncovered. Actually, \GS \cite{Cocchi2} and Terzan 2 \SAX best fits slightly favour a complete coverage of both the seeds.  This model is different from the double-Compton one proposed in \cite{Thompson} and \cite{Thompson2}: in this case there is just a single Compton region and the parameters space is similar to that of the traditional {\sc bb}+\CompTT model.

As a further development, simultaneous timing/spectral analysis can be helpful to better characterize these two LHS subclasses. In particular, 2P power spectra could be more affected by noise at the characteristic frequencies of the inner disk/TL because the hotter photon population of this compact region is supposedly brighter. 
\XTE-PCA and (future) {\em Astrosat} data can be effective for this kind of investigations.





\bibliographystyle{aipproc}   

\bibliography{sample}

\IfFileExists{\jobname.bbl}{}
 {\typeout{}
  \typeout{******************************************}
  \typeout{** Please run "bibtex \jobname" to optain}
  \typeout{** the bibliography and then re-run LaTeX}
  \typeout{** twice to fix the references!}
  \typeout{******************************************}
  \typeout{}
 }


\end{document}